\documentclass[a4paper,twoside]{article}
\newcommand{\affil}[1]{$^{\rm #1}$}
\usepackage[authoryear]{natbib}
\bibpunct{(}{)}{;}{a}{}{,}
\usepackage{graphicx}
\date{} 
\title{\large\bf\flushleft Type-Ia Supernova Rates and the Progenitor Problem\\ -- a Review}
\author{\parbox{\textwidth}{\flushleft
\vspace{-0.5cm}
{\it Dan Maoz\affil{A}, and Filippo Mannucci\affil{B,C}}\\
\vspace{0.4cm}
{\small \affil{A}\,School of Physics and Astronomy, Tel-Aviv
  University, Tel-Aviv 69978, Israel, maoz@astro.tau.ac.il}\\
{\small \affil{B}\,INAF - Osservatorio Astrofisico di Arcetri, Largo
  E. Fermi 5, 50125 Firenze, Italy, filippo@arcetri.astro.it}\\
{\small \affil{C}\,Harvard-Smithsonian 
Center for Astrophysics, 60 Garden Street, 
Cambridge, MA 02138, USA }}}
\begin{document}
\twocolumn[
\begin{changemargin}{.8cm}{.5cm}
\begin{minipage}{.9\textwidth}
\vspace{-1cm}
\maketitle
%
%
\small{\bf Abstract:}
The identity of the progenitor systems
 of type-Ia supernovae (SNe Ia) is
a major unsolved problem in astrophysics.
SN Ia rates are providing some striking clues.
We review the basics of SN rate measurement, 
preach about some sins of SN rate
measurement and analysis, and illustrate one of these 
sins with an analogy about Martian scientists. 
We review the 
recent progress in measuring SN Ia rates in various
environments and redshifts, and their use to
reconstruct the SN Ia delay time distribution (DTD) -- the SN
rate versus time that would follow a hypothetical 
brief burst of star formation. A good number of DTD measurements,
using a variety of methods, appear to be converging.
At delays $1<t<10$~Gyr, these measurements show
a similar, $\sim t^{-1}$, power-law shape.
The DTD peaks at the shortest delays probed, although there is still some
uncertainty regarding its precise shape at $t<1$~Gyr. 
At face value, this result supports
the idea of a double-degenerate 
progenitor origin for SNe~Ia. Single-degenerate progenitors
may still play a role in producing short-delay SNe Ia, or perhaps
all SNe Ia, if the red-giant donor channel is more efficient 
than found by most theoretical models. Apart from the DTD shape,
the DTD normalization enjoys
fairly good agreement (though perhaps some tension),
among the various measurements, with a Hubble-time-integrated DTD
value of about $2\pm 1$ SNe Ia 
per $1000~M_\odot$ (stellar mass formed with a low-mass-turnover IMF). 
The SN Ia
numbers predicted by binary population synthesis models are
at least several times lower than implied by the observed rates.
A recent attempt to
characterize the local WD binary population suggests that the WD
merger rate can explain the Galactic SN Ia rate, but only if
sub-Chandra mergers lead to SN Ia events.
We conclude by pointing to some future directions that should 
lead to progress in the field, including measurement of the bivariate
(delay and stretch) SN Ia response function .


\medskip
\medskip
\end{minipage}
\end{changemargin}
]
\small

\section{The SN Ia progenitor\\ problem}
\label{ss.intro}
Many aspects of type-Ia
supernovae (SNe Ia) are still
poorly understood (see, e.g., the
recent review by Howell 2011, and 
elsewhere in this Special Issue).
In particular, the identity of the progenitor 
systems of SNe Ia has not yet been established. This is something 
of an embarrassment, given 
the central role of SNe as
distance indicators for cosmology (e.g., Riess et al. 1998; Perlmutter
et al. 1999), as  synthesizers of heavy
elements (e.g. Wiersma et al. 2011), as
sources of kinetic energy in galaxy evolution processes (e.g. Powell
et al. 2011), 
and as accelerators of cosmic rays (e.g. Helder et al. 2009). 
Two main 
competing progenitor scenarios have been on the table for some time.
In the single degenerate (SD)
model (Whelan \& Iben 1974), a carbon-oxygen 
white dwarf (WD) grows in mass through accretion from
a non-degenerate stellar companion -- a main sequence star, a subgiant, 
a helium star, or a red giant -- until it approaches the
Chandrasekhar mass, ignites, and explodes in a thermonuclear runaway. 
The accretion can occur through Roche-lobe overflow of through a wind.
In the double degenerate (DD) scenario (Webbink 1984; Iben \& Tutukov 1984), 
two WDs merge after losing energy and 
angular momentum to gravitational waves. The merger outcome may be
a super-Chandra-mass 
object that ignites and explodes, or a situation in which 
the more massive WD tidally disrupts and accretes
the lower-mass object, approaches the Chandrasekhar limit, and
explodes.   
Although decades have passed since they were proposed, 
 neither the SD nor the DD models can yet be clearly favored 
observationally. Contrary to the situation for core-collapse SNe, where
a good number of progenitor stars have been identified in pre-explosion 
images (see Smartt 2009, for a review), no such progenitor has ever been
convincingly detected for a SN~Ia (Maoz \& Mannucci 2008; Li et al. 2011c; see 
Voss \& Nelemans 2008, Nelemans et al. 2008, for an ambiguous case). 

Both models, SD and DD, suffer from problems,
theoretical and observational. In terms of SD theory, it has long been 
recognized that the mass accretion rate on to the WD needs to be
within a narrow range, in order to attain stable hydrogen burning on the
surface, and mass growth toward the Chandrasekhar mass. 
Too-low an accretion rate will lead to explosive
ignition of the accreted hydrogen layer in a nova event, which likely
blows away more material from the WD than was gained (e.g. Townsley
\& Bildsten 2005, but see Starrfield et al. 2009; Zorotovic et
al. 2011). 
Too-high an
accretion rate will lead to a red-giant-like expansion of the accretor.
 The self-regulation of the accretion flow by a wind emerging from the 
accretor,
as conceived by Hachisu, Kato, \& Nomoto (e.g. Hachisu et al. 1999), has
thus long been considered to be an essential element of the SD model. 
Questions, however, have been raised as to whether 
the entire picture does not 
require too much fine tuning (e.g., Cassisi et al. 1998; Piersanti et
al. 2000; Shen \& Bildsten 2007; Woosley \& Kasen 2011).

The SD model faces additional obstacles when it comes to observational
searches for its signatures. Badenes et al. (2007) searched seven
young SN Ia remnants for the wind-blown cavities that would be
expected in the wind-regulation picture. Instead, in every case it
appeared the remnant is expanding into a constant-density ISM (but see
Williams et al. 2011, for an exception). Leonard
(2007) obtained deep spectroscopy in the late nebular phase of several SNe
Ia, in search of the trace amounts of H or He that would be expected
from the stellar winds. None was found. Prieto et al. (2008) have
pointed out the SNe Ia have been observed in galaxies with quite low 
metallicities. This may run counter to the expectations that, at low
enough metallicities, the optical depth of the wind would become
small, and the hence the wind-regulation mechanism would become
ineffective (Kobayashi \& Nomoto 2009). A positive point for the SD model
has been the variable NaD absorption that has been detected in the spectra
of a few SNe Ia (Patat et al. 2007; Simon et al. 2009) and has been
interpreted as circumstellar material from the companion.
However, it is unclear why
 such variable absorption is seen in only a minority of cases searched.
In a related development, Sternberg et al. (2011) have found some preference
for blue-shifted over red-shifted NaD absorption in single-epoch spectra
of 35 SNe Ia. They interpret the excess of blue-shifted absorptions 
as signatures
of the circumstellar material and conclude that $>20-25$\% of SNe Ia in spirals
would then derive from SD progenitors. Shen et al. (2011) have noted, however,
that such signatures could also arise in a post-merger, pre-explosion, wind  
in the DD scenario.

The companion, in an SD scenario, will survive the explosion, and is
likely to be identifiable by virtue of its anomalous velocity,
rotation, spectrum, or composition (e.g., Wang \& Han 2010). 
However, searches for the survivor 
of Tycho's SN have not been able to reach a consensus
(Ruiz Lapuente et al. 2004; Fuhrman 2005;
Ihara et al. 2007; Gonzalez-Hernandez et al. 2009; Kerzendorf et
al. 2009). Perhaps the effects of the explosion on the companion are
more benign than once thought (see Pakmor et al. 2008). Nonetheless, 
Hayden et
al. (2010), Bianco et al. (2011), Foley et al. 2012, and Ganeshalingam et al.
(2011), all
place observational limits on the presence
 of shock signatures of red-giant donors
in the light curves of SNe Ia with good
 early-time coverage, shocks that are expected from the ejecta hitting the
 companion, as calculated by Kasen (2010). Hancock et al. (2011) have used 
a stacking analysis of the VLA observations of Panagia et al. (2006), 
and Chomiuk et al. (2012) have 
used the EVLA, to set upper limits on the radio emission from 
SNe Ia in nearby galaxies. These limits challenge expectations if the SN
blastwave were encountering a circumstellar wind from the SD donor.

These same types of limits
were set more stringently than ever in the analysis of the recent SN
2011fe in M101, at 6.4~Mpc, which was discovered by the PTF survey 
less than a day after its explosion, and quickly followed up in many
wavebands.
 Li et al. (2011) used deep
pre-explosion images to rule out the presence of a red giant and
helium-star donors.
Horesh et al. (2012) set upper limits on both radio and X-ray emission,
excluding the presence of a circumstellar wind from a giant donor. 
Nugent et al (2011), Brown et al. (2011), and Bloom et al. (2011) used
very early optical and UV observations to exclude the presence of 
shocks from ejecta hitting a companion. They rule out red giants and,
in the latter two papers, also most main-sequence stars more massive
than the sun.

  Di Stefano (2010), and Gilfanov \&
Bogdan (2010), have both raised related arguments, 
that the accreting WDs in the SD scenario would be undergoing stable nuclear
burning on their surfaces, and hence would be visible as super-soft
X-ray sources (SSS), while the actual numbers of SSS are below those 
required to explain the observed SN Ia rate. Hachisu, Kato, \& 
Nomoto (2010) and Meng \& Yang (2011b) 
have countered that the theoretical SSS lifetimes and X-ray
luminosities have been overestimated in
this argument (see also Lipunov et al. 2011).

The DD model is also not free of problems. Foremost, it has long
been argued  that the merger of two unequal-mass WDs will lead to 
an accretion-induced collapse and the formation of a neutron star,
i.e. a core-collapse SN, rather than a SN Ia (Nomoto \& Iben 1985;
Guerrero et al. 2004; Darbha et al. 2011; Shen et al. 2011). 
Others, however, have proposed ways in which 
this outcome might be avoided (Piersanti et al. 2003; Pakmor et al. 2010; 
Van Kerkwijk et al. 2010; Guillochon et al. 2010; Shen et al. 2011).
Observationally, it has been much harder to find evidence either for
or against the DD scenario because, almost by construction, it leaves
essentially no traces. The most promising 
avenue has been to search the solar neighborhood for the close and
massive WD binaries that will merge within a Hubble time, surpassing
(perhaps) the Chandrasekhar mass and presumably 
producing DD SNe Ia.  
The largest survey to
date, SPY (Napiwotzki et al. 2004; Nelemans et al. 2005; 
Geier et al. 2007) has not found unambiguous super-Chandra merger progenitors
among $\sim 1000$ WDs, but an analysis of the results
that accounts for selection effects and efficiencies is still
lacking. Furthermore, a number of binary systems have been recently found
that may possibly evolve into super-Chandra, Hubble-time WD mergers (Geier et al. 2010; Rodriguez-Gil et al. 2010; Tovmassian et al. 2010).
The ongoing SWARMS survey by 
Badenes et al. (2009), is searching for close binaries among a
larger sample of  WDs in the Sloan Digital Sky Survey 
(SDSS; York et al. 2000), though 
with lower spectral resolution than SPY. We return to this subject 
in Section~\ref{wdbinaries}.

There are additional problems that are shared by both scenarios, SD
and DD. The energetics and spectra of the explosions do not come out
right, unless finely (and artificially) tuned in an initial subsonic
deflagration that, at the right point in time, spontaneously evolves into
a supersonic detonation (Khokhlov 1991). 
If the ignited mass is always near-Chandrasekhar, why is there the 
range of SN Ia luminosities inherent to the Phillips (1993) relation (see, 
e.g., Seitenzahl et al. 2011)?
Why is there a dependence of the SN Ia luminosity (or, equivalently, 
the mass of radioactive Ni synthesized) on the age of the galaxy host
 -- the oldest hosts, with little star formation, tend to
host faint, low-stretch, SNe Ia, while star-forming galaxies more likely host
bright-and-slow 
SNe Ia (e.g. Neill et al. 2009; Howell et al. 2009; Hicken et al. 2009; 
Sullivan et al. 2010). Finally, both scenarios
predict, based on binary population synthesis, SN rates that are lower
than actually observed (more on this later). 

Some variants of the SD and DD models have been conceived. The
near-Chandrasekhar-mass conjecture has come under renewed scrutiny.
Sub-Chandrasekhar explosions have been proposed as a way of explaining
some, or perhaps even most SN Ia events (Raskin et al. 2009; Rosswog et
al. 2009; Sim et al. 2010; van Kekwijk, Chang \& Justham
2010; Guillochon et al. 2010; Ruiter et al. 2011). 
Conversely, the Ni mass deduced for some SN Ia explosions is
strongly suggestive of a super-Chandrasekhar-mass progenitor
 (e.g. Tanaka et al. 2010; Silverman et al. 2010; Scalzo et al. 2010). 
Several ``SN on hold'' 
scenarios have also been proposed. Distefano et al. (2011) and Justham
(2011) have argued, in the context of the SD model, that a WD that had
grown to the Chandra mass could be rotation-supported against collapse
and ignition, potentially for a long time, during which the traces of
the messy accretion process (or even of the donor itself) would
disappear. 
Kashi and Soker (2011) propose a ``core-degenerate''
model, in which a WD and the core of an AGB star merge already in the
common-envelope phase. The merged core is supported by
rotation, again for potentially long times, until it slows down via
magnetic dipole radiation, and finally explodes (Ilkov \& Soker 2011).
Single-star SN Ia progenitor models have also been occasionally considered
(Iben \& Renzini 1983; Tout 2005), in which the degenerate carbon-oxygen
core of an AGB star is somehow ignited after it has lost its hydrogen
envelope (as it must, if the SN is to appear as a type-Ia, with no hydrogen 
in its spectrum). Waldman, Yungelson, \& Barkat (2008) have proposed a model
in which a binary companion is responsible for stripping the envelope off the
core, which then goes on to explode as a single star.   

In view of the above problems,
it has been realized for some time that measurement of SN Ia rates
may provide some critical discrimination among the various progenitor 
scenarios. In essence, finding the dependence of the SN rate on the 
age or age distribution of the 
host stellar population can reveal the age distribution of the
SN Ia progenitor population. Different progenitor scenarios involve
different timescales that control the production rate of SN Ia events.
Thus, SN rates can test progenitor models. 

\section{SN Ia rates}
  
Measuring a SN rate is, in principle, straightforward (but see
Section~\ref{parable}, below). One monitors a
sample of galaxies (a ``galaxy-targeted'' survey), 
or a region of sky to some depth, corresponding 
to some monitored volume (a ``field'' or ``volumetric'' survey). 
Nowadays, SNe are generally discovered via
image subtraction techniques, which, of course, turn up all sorts of
SNe, plus other contaminants, both real (e.g. LBV ``impostors'', 
active galactic nuclei,
variable Galactic stars, Solar-system objects), and artificial (cosmic
ray events, imperfect subtraction residuals). SN surveying need not 
necessarily be based on imaging, as in the searches for SNe in 
SDSS 
galaxy spectra by Madgwick et al. (2003) and  Krughoff et al. (2011),
and to which all that follows below applies equally well as for imaging.   

After identifying the real SNe and their types (often not an easy
task), the SN rate, e.g. in a galaxy-targeted survey, will be
\begin{equation}
R=\frac{N_{\rm SN}}{\sum\limits_{i} t_i},
\label{snrate1}
\end{equation}
where $ N_{\rm SN}$ is the number of (say) SNe Ia discovered, $t_i$
is the ``control time'' or ``effective visibility time'' of each
galaxy in the survey, and the sum is over all galaxies that were
monitored. The visibility time is the time during which a SN of the
given type could have been detected during the survey. 
In a ``rolling survey'', a sample is monitored continuously with 
cadences that are shorter than the rise and fall time of the targeted
SNe, and the observations are deep enough to catch the targeted 
SNe at least during their maximum light. In that case, the visibility
time is simply the duration of the survey (or the sum of various
seasons during which it was undertaken). In ``one-shot'' surveys,
where cadences are much longer than SN variation times, the visibility
time of a galaxy is, in principle,
 the time during which a SN would be above the flux
limit. 

In practice, there are never ``on-off'' flux limits, but rather  
detection efficiencies as a function of SN magnitude. The visibility
time calculation that accounts for this (and hence the addition of
``effective'' to ``visibility time'') is
\begin{equation}
t=\int_0^\infty \epsilon(m) \left(\frac{dm}{dt}\right)^{-1} dm ,
\end{equation}
where $m(t)$ is the light curve of the targeted SNe (in the rest-frame
band that corresponds to the observed band of the survey), and $\epsilon (m)$
is the detection efficiency as a function of magnitude $m$. In real 
situations, the detection efficiency will often depend on the stellar 
background -- SNe will be harder to detect the closer they are to the
centers of their hosts, and the higher is the surface brightness of 
 those hosts. To deal
with those realities, the only reliable way of estimating detection
efficiency is through simulations: many fake SNe are planted at random 
in the real data, but at positions that track the stellar light\footnote{In reality, the degree to which the SNe track the stellar light will vary depending on 
the photometric band (e.g., Raskin et al. 2009), but for the purpose
of simply estimating detection efficiency versus SN magnitude, 
this is a suitable approximation.}, and 
recovered via the same process used for the real SNe. The recovered
fraction gives $\epsilon(m)$. Since different images in a survey will
generally have different observing conditions (depth, seeing, etc.),
ideally the efficiency curve should be determined for every image.
SNe Ia, and certainly other types of SNe, have a diversity of light
curve shapes, with (for SNe Ia) 
correlated peak luminosities. This diversity needs
to be taken into account when calculating the visibility time, by
drawing light curves $m(t)$ from their intrinsic distributions. This
last point is complicated by the fact that the intrinsic SN luminosity 
functions (e.g., Li et al. 2011a) 
are poorly known -- the measured functions contain, to some
degree, the flux limits and selection effects of the surveys in which 
the SNe were discovered, and the host galaxy extinctions, which vary with 
host population and with redshift. Thus, a visibility time calculation 
needs to assume an intrinsic SN luminosity function and a particular 
distribution of extinctions. These assumptions will propagate into the  
final derived SN rate. The uncertainties regarding these assumptions
will translate into systematic rate uncetainties. 

SN rates are most interesting when expressed as rest-frame quantities,
and therefore, in rolling surveys, the visibility time is reduced by 
$(1+z)^{-1}$. In one-shot surveys, however, the lower rate at which
the SNe appear to go off at high $z$, due to this cosmological time dilation
(and which, alone, would lead
to a smaller number of
detected SNe), is cancelled by the slower
apparent evolution of each SN (which makes the SN detectable for a longer
time.) Hence, the number of SNe detected in one-shot surveys is
unaffected by time dilation, and using the observer-frame visibility time in
Equation~\ref{snrate1} gives the rest-frame SN rate.

The rate given by Equation~\ref{snrate1}, as it stands, is not of much
use. For example, in a galaxy-targeted survey it would give the SN rate per
average galaxy in the survey. To be physically useful, a SN rate needs
to be normalized by some property relating to the monitored
population. In rates from field surveys, which started in earnest with
the cosmological SN surveys of the mid 1990's, the normalization is
by unit comoving volume. In galaxy-targeted surveys, the convention, until
recently, was to normalize the rate to a unit stellar luminosity in
some photometric band (often $B$). However, luminosity, especially
$B$-band luminosity, is more a tracer of star-formation rate than of
stellar mass, and is a rapidly varying function of stellar age. Mannucci et
al. (2005) introduced the normalization of SN rates by stellar mass, 
with interesting consequences, as we will see below.
A mass-normalized SN rate will be 
\begin{equation}
R=\frac{N_{\rm SN}}{\sum\limits_{i} M_i t_i},
\label{snrate2}
\end{equation}
where $M_i$ is the stellar mass of the $i$th galaxy in the survey.

As already mentioned, 
SN surveys can be divided based on their targeting
scheme (with some surveys being combinations of several schemes): surveys
targeting specific galaxies; field surveys, that monitor some volume of space;
and also surveys targeting specific galaxy clusters. Until recently, the
best local-galaxy-targeted SN sample was the one defined by Cappellaro et
al. (1999), based on a compilation of several visual and photographic
surveys. Rates based on this sample were derived most recently by
Mannnucci et al. (2005). SN Rates from a new
survey of nearby southern-hemisphere
galaxies have been presented by Hakobyan et al. (2011).
The Lick Observatory SN Search (LOSS),
conducted over the past 15 years, is now
the largest survey for
local ($<200$~Mpc) SNe . It has
produced a homogeneous set of over 1000 SNe (274 of them SNe Ia)
detected via
 CCD surveying with the robotic KAIT telecope. The survey 
and the resulting SN rates have been presented  in Li et al. (2011a,b),
Leaman et al. (2011), and Maoz et al. (2011). 
A recent compilation of rates based on field (rather than
galaxy-targeted) surveys
is included in
Graur et al. (2011, see also Section~\ref{snrvsz}, below). 
Galaxy cluster SN Ia rates have been compiled in 
Maoz et al. (2010, see also Section~\ref{clustersnr}, below).  

 The vast majority of known SNe Ia have been discovered in surveys
at optical wavelengths, that enjoy large-area detectors and low 
sky brightness. However these surveys miss the highly 
extinguished SNe that are known to occur in star forming galaxies (di Paola 
et al. 2002). Near-IR SN surveys focused on star-forming galaxies
have indeed yielded extinguished SNe, both
 core-collapse and SNe Ia (e.g., 
Maiolino et al. 2002;
Mannucci et al. 2003; 
Mattila et al. 2007;
Cresci et al. 2007;
Kankare al. 2008)

\section{The delay-time distribution}
\subsection{The theoretical DTD}
\label{theory}
A fundamental function that can shed light on the progenitor question is    
the SN delay time distribution (DTD). The DTD is the hypothetical SN
rate versus time that would follow a brief burst of star formation,
which formed a unit total mass in stars. 
In other contexts, the DTD would be called the transfer function, the
response function, the Green's function, the kernel, the point-spread 
function, and so on, that characterizes the system. It is the 
``impulse response'' that embodies the
physical information of the system, free of nuisances such as, in the present
context, e.g., the star-formation histories (SFHs) 
of the galaxies hosting the SNe.
The DTD is directly linked to the lifetimes (hence,
the initial masses) of the progenitors and to the 
binary evolution timescales up to the
explosion, and therefore different progenitor scenarios predict
different DTDs. The DTD could conceivably vary with environment 
or cosmic time, due to, e.g., changes in initial mass function (IMF) or
in metallicity, but for the moment we will ignore this complication.  

Various theoretical 
forms have been proposed for the DTD. Some have been derived from detailed
``binary population synthesis'' calculations, where one begins with a large
population of binaries with a chosen distribution of 
initial parameters, and one models  the various stages 
of their stellar and binary evolution, including mass loss, mass transfer, and
common-envelope evolution (with its physics parametrized in some way)
(e.g., Han et al. 1995; Jorgensen et al. 1997; Yungelson \& Livio 2000;  
Nelemans et al. 2001; Han \& Podsiadlowski
2004; Lipunov et al. 2009; Ruiter et al. 2009, 2011;
Mennekens et al. 2010; Wang, Li, \& Han 2010; 
Meng et al. 2011; Bogomazov \& Tutukov 2009, 2011). 
Other theoretical DTDs have been
based on physically motivated mathematical parameterizations, with varying
degrees of sophistication (e.g., Greggio \& Renzini 1983; 
Tornambe \& Matteucci 1986;
Ciotti et al. 1991; Sadat et
 al. 1998; Madau
et al. 1998; Greggio 2005, 2010; Totani et al. 2008). Finally, 
some DTDs have been ad hoc formulations intended to
reproduce the observed field-SN rate evolution (e.g., Strolger et
al. 2004, 2010).

Some generic features of the DTD for the DD and SD models
can be derived from simple
physical considerations, and generally emerge also in the more detailed
models.
 As noted by previous authors (e.g., Greggio 2005;
Totani et al. 2008) a power-law DTD time dependence
is generic to models (such as the DD model) in which
the event rate ultimately depends on the loss of energy and angular
momentum
to gravitational radiation by the progenitor binary system.
If the dynamics are controlled solely by gravitational wave losses,
the time $t$ until a merger depends on the
binary separation $a$ as
\begin{equation}
t\sim a^4,
\end{equation}
with a weaker dependence on the WD masses, which in any case 
are in a limited range.
If the initial separations are distributed as a power law
\begin{equation}
\frac{dN}{da}\sim a^\epsilon,
\end{equation}
then the event rate will be
\begin{equation}
\frac{dN}{dt}=\frac{dN}{da}\frac{da}{dt}\sim t^{(\epsilon -3)/4} .
\label{DDdependence}
\end{equation}
For  a fairly large range around $\epsilon\approx -1$, which describes
well the observed distribution of initial separations  of
non-interacting
binaries (see Maoz 2008 for a review of the issue in the present
context),
the DTD will
have a power-law dependence with index not far from $-1$.
Indeed, as noted, 
a $\sim t^{-1}$ power law appears to be a generic outcome also of 
detailed binary population synthesis calculations of the DD channel
(e.g., Yungelson \& Livio 2000;   
Mennekens et al. 2010).
Of course, in
reality, the binary separation distribution of WDs that have
emerged from their common envelope phase could be radically different,
given the complexity of the physics of that phase, and need not even follow
a power law. Thus, the $\sim            
t^{-1}$ DTD dependence of the DD channel cannot be considered
unavoidable (see e.g. Ruiter et al. 2009). 
Nevertheless, a post-common-envelope separation
distribution that is about flat in log separation (i.e., $\epsilon=-1$)
does seem to emerge from many
simulations (e.g., Claeys et al., in preparation).

A different power-law DTD dependence, with different physical
 motivation, has
been proposed by Pritchet et al. (2008), by way of interpreting
 volumetric SN rates in the SNLS (but see Greggio 2010). 
If the time between formation
of a WD and its explosion as a SN~Ia is always brief compared to the
formation time of the WD, the DTD will simply be proportional to the
formation rate of WDs. Assuming that the main-sequence lifetime of a
 star
depends on its initial mass, $m$, as  a power law,
\begin{equation}
t\sim m^\delta,
\end{equation}
and assuming the IMF is also a power law,
\begin{equation}
\frac{dN}{dm}\sim m^\lambda,
\end{equation}
then the WD formation rate, and hence the DTD, will be
\begin{equation}
\frac{dN}{dt}=\frac{dN}{dm}\frac{dm}{dt}\sim
t^{(1+\lambda-\delta)/\delta} .
\end{equation}
 For the commonly used value of $\delta=-2.5$ (from stellar evolution
 models) and the Salpeter
 (1955) slope of $\lambda=-2.35$, the resulting power-law index is
 $-0.46$, or roughly $-1/2$. Pritchet et al. (2008) raised the 
possibility
of such a $t^{-1/2}$ DTD.  
It is arguable that, instead of
 a single, $\sim t^{-1}$ power law, motivated by binary mergers,
with this power law
  extending back to delays as short as
 40~Myr (the lifetime of the most massive stars that form WDs), 
there could be a ``bottleneck'' in the supply of progenitor
systems below some delay. Such a bottleneck could be
 due to the birth rate of WDs, which
behaves as $\sim t^{-1/2}$. One possible result would then be a
  DTD, $\Psi(t)$, that is a broken-power-law, 
with $\Psi\propto t^{-1/2}$ up to some
  time, $t_c$, and $\Psi\propto t^{-1}$ thereafter.
A possible value could be  $t_c\approx 400$~Myr, corresponding to the
lifetimes of $3M_\odot$ stars. If that were the lowest initial mass of
 stars that can produce the WD secondary in a DD SN~Ia progenitor, then
beyond $t_c$ the supply of new systems would go to zero, and the
SN~Ia rate would be dictated by the merger rate. For example, the Greggio
 (2005) DD-wide model is indeed a
 $t^{-1/2}$,~$t^{-1.3}$, broken power-law with break at
$t_c<400$~Myr. In sub-Chandra merger SN Ia models (Sim et al. 2010;
 Van Kerkwijk et al. 2010; Ruiter et al. 2011), involving the mergers
of white dwarfs that had main sequence masses smaller than $3{\rm
 M_\odot}$, $t_c$ would shift to longer delays .

In contrast to the DD model, for the SD model there is a
large variety of results among the predictions for the DTD. 
Some of this variety is due to the fact that ``SD''
includes an assortment of very different sub-channels. Some of it is
due to the fact that, even within a given sub-channel, different
workers treat the same evolutionary phases using different
approximations
 (e.g. the 
common-envelope phase phase, via the Webbink (1984) 
$\alpha$ formalism, or the Nelemans \& Tout (2005)
$\gamma$ parameter). And some of of the variety is due the use of
different assumed input parameters and distributions. But,
disturbingly, attempts by some teams (e.g. Mennekens et al. 2010) to reproduce
results of other teams by using the same recipes and inputs still show
significant discrepancies. Under this state of affairs, it may be
that the theoretical SD predictions for the DTD have not yet reached the 
point where they can be meaningfully compared to the observations. 
 
However, one generic
prediction that SD models often do seem to make is that the DTD
tends to drop off sharply after a few Gyr, which can be understood as follows.
The timescale of the mass transfer phase is only of order
millions of years, much less than the other timescales in the problem.
In SD systems where the donor is a main-sequence star, the timescale
for explosion is therefore dictated by the time required for magnetic 
braking that reduces the separation, leading
 the donor to fill its Roche lobe. In systems 
where the donor is a subgiant star that has just evolved off the main sequence,
the dominant timescale is the donor's evolutionary timescale.
As we progress down the stellar mass function 
to lower and lower primary masses, we produce lower and lower mass
WDs. These, in turn, require larger and larger mass transfers from the 
companion to make up the mass difference required for the WD 
to reach near the Chandrasekhar mass. Donors with too-low masses 
cannot transfer 
material at the necessary quantities and rates (Greggio 2005).
After a few Gyr, the 
secondaries are not massive enough for the job, and 
 the DTD drops.

A caveat to this description may be the existence of a ``symbiotic'' SN Ia
SD channel, in which the donor star is a red giant. In the version  
worked out by Hachisu, Kato, \& Nomoto (2008a), mass 
is stripped from the giant by the wind from the WD, and accreted 
onto the WD. This permits 
large mass transfers to the WD from relatively low-mass secondaries,
down to $\sim 0.9 M_\odot$, 
producing SNe Ia also at very large delays. 
Blondin et al. (2010) find a low stripping efficiency in hydrodynamical 
simulations of the process. However, 
an alternative, tidally enhanced, 
rather than stripped, donor wind has been proposed
by Chen, Han,\& Tout (2011).
Hachisu, Kato, \& Nomoto (2008b)
show that the DTDs from their SD main-sequence and red-giant
channels can combine to give 
a $t^{-1}$ DTD out to long delays, similar to the 
DTD described above for the DD scenario. 
However, most other binary
population synthesis models find that the red-giant SD channel is
highly inefficient, and will contribute negligibly to the DTD. For 
example, Han \& Podsiadlowski
(2004) and Wang et al. (2010) 
find that the red-giant SD channel  contribution to the SN Ia rate
is 30-60 times lower than that of the main-sequence SD channel.  

Keeping this possible caveat in mind, it appears that two generic
DTD expectations that we can remember
from the two main models are: for DD, a roughly 
$t^{-1}$ dependence, at least beyond $\sim 1$~Gyr, and extending out to 
a Hubble time; and for SD, a cutoff in the DTD beyond a few Gyr.

\subsection{The observed DTD}
 Until recently, only few, and
often-contradictory, observational constraints on the DTD existed. In
the
past few years the observational situation has changed
dramatically. A range of different approaches to recover the DTD,
using a variety of SN samples, environments, and redshifts,
are yielding a 
 consistent view of the DTD, one that is beginning to
discriminate among the SN Ia progenitor models.
We review these observations, with emphasis on the most recent ones.

\subsubsection{SN Ia rates versus redshift\\ in galaxy clusters}
\label{clustersnr}
We will start with a method 
for recovering the DTD that is, conceptually, perhaps the most simple
to grasp -- by measuring the SN rate
vs. redshift in massive galaxy clusters. As explained below,
the deep potential wells of clusters, combined with their
relatively simple SFHs, 
make them ideal locations for
measuring the DTD.
Optical spectroscopy and multiwavelength photometry 
of cluster galaxies has shown consistently that the bulk of their
stars were formed within short episodes ($\sim 100$~Myr) 
at $z\sim 3$ (e.g., Daddi et al. 2000; Saracco et al. 2003; Stanford et al. 2005; van Dokkum and van der Marel 2007; Jimenez et al. 2007;
 Eisenhardt et al. 2008). Thus, the
observed SN Ia rate vs. cosmic time $t$, given  
a stellar formation epoch $t_f$,  
provides an almost direct
measurement of the form of the DTD,
\begin{equation}
R_{Ia}(t)=\frac{\Psi(t-t_f)}{m(t-t_f)}.
\label{ddtmassloss}
\end{equation}
Here, $m(\tau)$ is the surviving mass fraction in a stellar
population, after accounting for the mass losses during stellar
evolution due to SNe and winds, and is obtainable from stellar population
synthesis models.
Here and throughout, we will be
considering SN rates measured per unit stellar mass {\it at the time of
observation}, and DTDs normalized per unit stellar mass {\it
  formed}. In making intercomparisons of measurements among
themselves, and with predictions, it is important that consistent
definitions and  stellar IMFs be assumed (see
Section~\ref{parable}).

Furthermore, the record of metals trapped in stars and in    
the intracluster medium (ICM) by the cluster gravity 
constrains the  integrated 
number of SNe Ia per formed stellar mass, $N_{\rm SN}/M_*$, that have
exploded in the cluster over its stellar age, $t_0$, 
and hence the normalization of the DTD,
\begin{equation}
\int_0^{t_0} \Psi (t) dt=\frac{N_{\rm SN}}{M_*}.
\label{normconstraint}
\end{equation} 
As reviewed in detail in Maoz et al. (2010), X-ray and optical
observations of galaxy clusters have reached the point where they 
constrain $N_{\rm SN}/M_*$ to the level of $\pm 50\%$, based
on the observed abundances of iron (the main product of SN Ia
explosions), after accounting for the contributions by core-collapse
SNe (and the uncertainty in that contribution).
 \begin{figure}
  \includegraphics[height=.25\textheight]{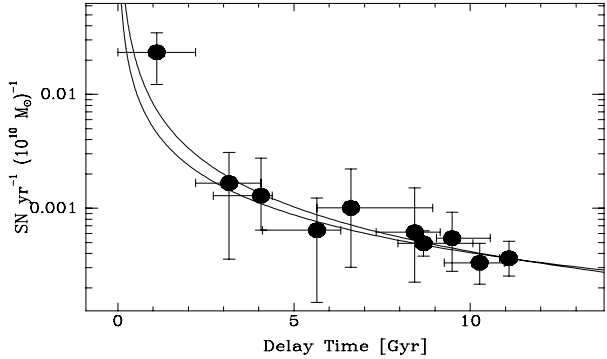}
  \caption{Points: 
SN Ia DTD recovered based on galaxy cluster SN Ia rate measurements,
and cluster iron abundances, from Maoz et al. (2010). Vertical axis
DTD values, here and throughout the paper, are per formed stellar 
mass, assuming a ``diet Salpeter'' IMF (Bell et al. 2003), that simulates
the effect of a realistic IMF with a low-mass turnover. 
The solid
curves are power laws, $t^{-1.1}$ and $t^{-1.3}$,
 that describe well these results.
}
\label{clusterratefig}
\end{figure}

A decade ago, there were no real measurements of SN rates in galaxy clusters. 
However, 
the observational situation has improved dramatically,
especially in the last few years. Following large investments of
effort
and observational resources,
fairly accurate  cluster SN~Ia rates have now been measured 
in the redshift range
$0<z<2$ (Gal-Yam et al. 2002, 2008; Sharon et al. 2007, 2010;
Mannucci et al. 2008; Graham et al. 2008; Dilday et al. 2010b; Barbary
et al. 2012a; Sand et al. 2011). 
Figure~\ref{clusterratefig} shows the DTD derived by Maoz
et al. (2010) based on most of these galaxy-cluster SN Ia rate
measurements, together with the iron-based DTD integral constraint,
which sets the level in the earliest DTD bin. Note the excellent
agreement with a $\sim t^{-1}$ form. 

A possible caveat to this picture is that galaxy clusters consist,
in addition to early-type galaxies, 
also of spiral galaxies, which have ongoing star formation. Furthermore,
even 
early-type galaxies sometimes show traces of recent 
star formation, as evidenced
in local ellipticals by, 
e.g., dust features (e.g. Colbert et al. 2001), cold molecular gas (e.g. Young et al. 2009; Temi et al. 2009), or
blue UV colors (e.g. Kaviraj et al. 2010; Rampazzo et al. 2011; see Schiavon 
2010 for a review).
In principle, this deviation 
from the assumption of a brief, high-$z$, burst of star formation, could affect
the derived DTD, as some of the SNe Ia observed in any cluster sample would
be due to these younger progenitors. In practice, however, several lines of 
evidence suggest this may not
be a serious problem. As discussed by Maoz et al. (2010), 
most of the cluster surveys that produced the rates shown above
have monitored only the central regions, at radii of order $R< 500$~kpc, 
which are completely 
dominated by early-type, rather than spiral, galaxies. Indeed, the
majority of the SNe Ia that these surveys
 have discovered have been hosted by ellipticals.
In terms of ongoing star formation in the early-types,
Maoz et al. (2010) have shown that  
the $t^{-1}$ conclusion  is weakly
dependent on the various assumptions laid out
above, such as the precise redshift of cluster star formation, whether
it was a brief or extended burst, or the contribution of ongoing
low-level star formation in clusters, as long as these are 
at the levels, redshifts, and
cluster locations allowed by direct measurements of  
star formation tracers in clusters.

\subsubsection{SN Ia rates versus redshift, compared to cosmic
  star-formation history}
\label{snrvsz}
Another observational 
approach to recovering the DTD 
has been to compare the volumetric SN rate from field surveys, as a
function of redshift, to the cosmic SFH. Given that
the DTD is the SN ``response'' to a short burst of star formation, 
the volumetric SN rate versus cosmic time, $R_{Ia}(t)$, 
will be the convolution of the DTD with the SFH (i.e. the 
star formation rate per unit comoving volume versus cosmic time, $\dot
\rho(t)$),
\begin{equation}
R_{Ia}(t)\propto\int_{0}^{t}\dot\rho(t-\tau)\frac{\Psi(\tau)}{m(\tau)}d\tau
,
\end{equation}
where $m(\tau)$ is again the surviving mass fraction in a stellar
population.

Gal-Yam \& Maoz (2004) carried
out the first such comparison, using a small sample of SNe~Ia out to
$z=0.8$, and concluded that the results were strongly dependent on the
poorly known cosmic SFH, 
a conclusion echoed by Forster et al. (2006).
 With the availability of SN rate measurements
to higher redshifts, Barris \& Tonry (2006) found a SN~Ia rate that
closely tracks the SFH out to $z\sim 1$, and concluded  that the DTD
must be concentrated at short delays, $<1$~Gyr. Similar
conclusions have been reached, at least out to $z\sim 0.7$, by
Sullivan et al. (2006) and Mannucci, Della Valle, \& Panagia
 (2006). In contrast,
Dahlen et al. (2004, 2008) and Strolger et al. (2004, 2010) 
have argued for a DTD
that is peaked at a delay of $\sim 3$~Gyr, with little power at short
delays, based on a sharp decrease in the SN~Ia rate at $z>1$ found by
them in the Hubble Space Telescope (HST) GOODS survey. However,
Kuznetzova et al. (2007) re-analyzed some of these datasets and
concluded that the small numbers of SNe and their potential
classification  errors preclude reaching a conclusion. Similarly,
Poznanski et al. (2007) performed new measurements of the $z>1$ SN~Ia
rate by surveying the Subaru Deeep Field with the Subaru Telescope's
SuprimeCam. They
 found that, within uncertainties, the SN rate could be
tracking the SFH. This, again, would imply a short delay time.
Mannucci et al. (2007) and 
Greggio et al. (2008) pointed out that underestimated extinction
of the highest-$z$ SNe, observed in their rest-frame ultraviolet
emission, could be an additional factor affecting these results.
Blanc \& Greggio (2008) and Horiuchi \& Beacom (2010)  have shown that, within the errors, a wide
range of DTDs is consistent with the data, but with a preference for
a DTD similar to $\sim t^{-1}$. 
\begin{figure}
  \includegraphics[height=.25\textheight]{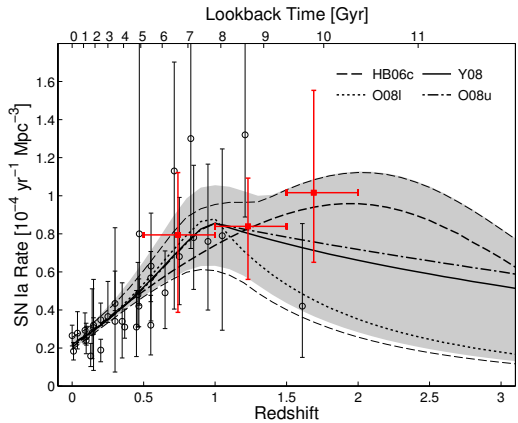}
  \caption{Compilation of volumetric 
SN Ia rates versus redshift (see Graur et al. 2011,
for references). Filled squares (red) 
are from the Subaru
  Deep Field search by Graur et al. (2011). The various curves are
  obtained by convolving various SFHs with the best-fit
  DTD, which in all cases has a form of approximately 
$\Psi(t)\propto t^{-1}$. The
shaded area is the combined 68\% confidence region resulting from
the statistical uncertainties in the rates, and the different possible SFHs.
}
\label{fieldrates}
\end{figure}

Happily, it appears that the picture
is finally clarifying and converging with respect to the field SN Ia rate
as a function of redshift, and the DTD that it implies. 
Rodney \& Tonry (2010) have presented a 
re-analysis of the data of Barris \& Tonry (2006), with new SN Ia
rates that are lowered, and in much better agreement with other measurements 
at similar redshifts. Accurate new rates from the Supernova Legacy 
Survey (SNLS; Gonz\'alez-Gait\'an et al. 2011; Perrett et al., in
preparation;
see also Kistler et al. 2011) 
agree with the revised numbers, and suggest a SN Ia rate
that continues to rise out to $z=1$, albeit growing more gradually
than the SFH. Finally, a quadrupling of the initial Subaru Deep Field
high-$z$ SN sample,
first presented by Poznanski et al. (2007), is resolving the puzzle of the
SN rate out to $z=2$. Graur et al. (2011) present a sample
of 150 SNe discovered by ``staring'' at this single 
field at four independent  epochs, 
with 2 full nights of integration per epoch. SN host galaxy redshifts
are based on spectral and photometric redshifts, from the
extensive
UV to IR database existing for this field. Classification of the SN
candidates is photometric. The SN sample includes 26 events that are 
fully consistent with being normal SNe Ia in the redshift range
$1.0<z<1.5$, and 10-12 such events at $1.5<z<2.0$. 
The rates derived from the Subaru data, now based on much better statistics
than the GOODS results, merge
 smoothly with the most recent and 
most accurate rate measurements at $z<1$, confirming the trend of
a SN Ia rate that gradually levels off at high $z$, but does not dive
down, as previously claimed by Dahlen et al. (2004, 2008).
Graur et al. (2011) find that a DTD with a
power-law form, $\Psi(t)\propto t^{-1}$, when
convolved with a wide range of plausible SFHs, gives an excellent 
fit to the observed SN rates. Their formal result for the power-law index is 
$\beta=-1.1\pm 0.1$ (random error, due to the uncertainties in the SN rates),
$\pm 0.17$ (systematic error, due to the range of possible SFHs).
This conclusion is further confirmed when 
the Perrett et al. SNLS rates are also included 
in the fits (Kistler et al. 2011). 
The field-survey volumetric SN rates and fits are shown 
in Figure~\ref{fieldrates}. In addition to the works already mentioned,
this includes SN rates from Cappellaro
et al. (1999),
Hardin et al. (2000), Pain et al. (2002), Madgwick
et al. (2003), Tonry et al. (2003), Blanc et al. (2004), Neill
et al. (2006, 2007), Botticella et al. (2008), Dilday et al. (2008, 2010a), 
Horesh et al. (2008), and Li et al. (2011b). Additional high-$z$ rates
have been recently presented by Barbary et al. (2012b), and are
consistent with this picture.

\subsubsection{SN Ia rate versus galaxy ``age''}
\begin{figure}
  \includegraphics[height=.34\textheight]{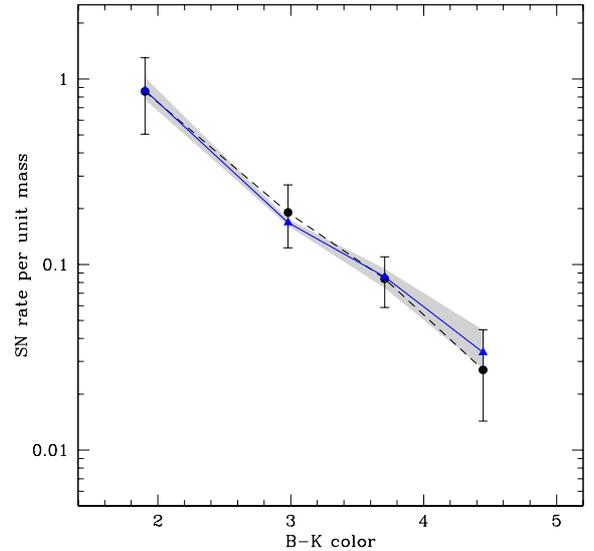}
  \caption{Observed SN Ia rate per unit stellar mass (circles with error bars),
 as a function
of galaxy $B-K$ color, from Mannuci et al. (2005). Rates are in units of
$10^{-12}~{\rm yr}^{-1}~M_\odot^{-1}$. Here the mass is the existing 
stellar mass at the time of observation, assuming the Bell et al. (2003)
diet-Salpeter IMF.
Triangles are model predictions
based on a $\sim t^{-1}$ DTD. In the model,
 each galaxy color
corresponds to an exponential SFH with some characteristic timescale,
such that the observed present-day color is reproduced. 
Each SFH, when convolved 
with a $t^{-1.1}$ DTD, reproduces the observed rates. 
The shaded area is the uncertainty in the predictions 
due to the uncertainty in the galaxy stellar populations 
(see Mannucci et al. 2006, for details).
 }
\label{snmcolor}
\end{figure}

Another approach to recovering the DTD has been to compare the SN
rates in galaxy populations of different characteristic ages. 
It is this approach that gave the first clear indications
for a range of delay times in the DTD. Mannucci
et al. (2005, 2006), analyzing the Cappellaro et al. (1999) SN sample,
discovered that the SN Ia rate per unit stellar mass depends on 
host galaxy parameters that trace the star-formation rate, such as
Hubble type or color. On the other hand, early-type galaxies with no
current star formation still have a non-zero SN Ia rate. 
This observation is shown in Figure~\ref{snmcolor}.
The dependence of SN rate on host color  
was confirmed by Sullivan et al.
(2006) for the SNLS sample as well. Both groups interpreted this
result
to indicate
 the co-existence of two SN~Ia
populations, a ``prompt'' population that explodes within  $\sim
100-500$~Myr, and a delayed channel that produces SNe~Ia on timescales of
order 5~Gyr. This led to the ``$A+B$'' formulation 
(Mannucci et al. 2005; Scannapieco \& Bildsten 2005), 
in which the SN Ia rate in a galaxy
is proportional to
both the star-formation rate in the galaxy (through the $B$ parameter, 
or through the core-collapse SN rate in Mannucci et al. 2005) 
and to the stellar mass of the galaxy (through the $A$ parameter). 

\begin{figure}
  \includegraphics[height=.25\textheight]{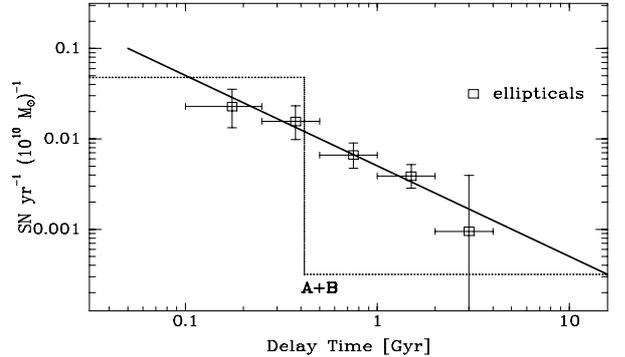}
  \caption{The two-bin DTD implied by the measured values of
the $A+B$ model, and the DTD recovered by Totani et al. (2008) 
by
comparing SN~Ia rates in early-type galaxies of different
characteristic ages, seen at $z=0.4-1.2$ in the SXDS.
The solid line shows the good agreement with a $t^{-1}$ power law.
 }
\label{ApBtotani}
\end{figure}

In 
essence, however, $A+B$ is just a DTD with two coarse time bins. 
The $B$ parameter,
divided by the assumed duration of the prompt component, is the mean
SN rate in the first, prompt, time bin of the DTD. The $A$ parameter
(after correcting for stellar mass loss, $m(t)$, of an old
population, always about a factor of 2), is
the mean rate in the second, delayed, time bin.
In retrospect, these two ``channels'' appear to be just 
integrals over a continuous DTD on two sides of some time border
(Greggio et al. 2008). And, the prompt and delayed SN Ia rates
corresponding to $A$ and $B$ define the logarithmic slope and normalization 
of a power law. Because of the broad range of the time interval over which 
the DTD is effectively averaged to yield the $A$ parameter, this parameter
is really just a rough approximation to the mean of the DTD in this range, 
a mean that will vary among populations with diverse SFHs.
Nevertheless, a $t^{-1}$ power law is roughly consistent with the measured
values of $A$ and $B$, as seen in Fig.~\ref{ApBtotani}, where $A$ and $B$
are the medians of the values compiled by Maoz (2008).  

The directly observed dependence of the SN Ia rates on host galaxy
color, as seen in Fig.~\ref{snmcolor},
can be very well reproduced by a model that assumes a $\sim t^{-1}$
DTD (the same was shown by Greggio 2005, for some of her DD models).
In the model results shown here, each galaxy color
corresponds to an exponential SFH with some characteristic timescale,
such that the observed present-day color is reproduced. 
Each SFH, when convolved 
with a  $t^{-1.1}$ DTD, reproduces the observed rates 
(see Mannucci et al. 2006 for details).
  
Totani et al. (2008) used a similar approach to recover the DTD, by
comparing SN~Ia rates in early-type galaxies of different
characteristic ages, seen at $z=0.4-1.2$ as
part of the Subaru/XMM-Newton Deep Survey (SXDS) project.
They were the first to show observationally that the  DTD is 
consistent with a $t^{-1}$ form. The Totani DTD is also shown
is Fig.~\ref{ApBtotani}. 

 Additional 
recent attempts to address the issue with the ``rate vs. age''
approach have been made by Aubourg et al. (2008), Raskin et al. 
(2009), Yasuda \& Fukugita
(2009), Cooper et al. (2009), Schawinski (2009), and Thomson \& Chary (2011). 
They have generally confirmed the
existence of ``prompt'' SNe Ia, although with quite a wide range
in defining the age of that population. 
Furthermore, some of these studies
have compared, a posteriori, the properties of galaxies that were seen
to host SNe, to the properties 
of matched ``control samples'' of other galaxies. The risks 
of such a procedure are discussed in Section~\ref{parable}. 

While the concept of a typical age for a host galaxy, interpreted as 
a SN Ia progenitor age (e.g. Totani et al. 2008), 
has been useful, it is nonetheless only 
a rough (and often risky) zeroeth-order 
approximation to the full SFH of a galaxy.
The average SN Ia rate from a 
stellar population is not the same as the SN Ia 
rate of the average stellar population.
Mannucci (2009) has shown some concrete examples of how 
galaxies with 
similar mean ages, but
with different age distributions, can have SN rates that differ by 
orders of magnitude.
For example, as little as 0.3\%, by mass, 
of young ($10^8$~yr) stars that are added to an old 
($10^{10}$~ yr) galaxy
can easily boost its SN Ia rate by a factor of two. 
The galaxy remains old-looking, the mass-weighted mean 
age does not change much, but the observed rate is not due 
to the DTD at that delay. 
A DTD recovery method that avoids
this approximation is described next.

\subsubsection{SN Ia rate versus individual galaxy star formation histories}
Both of the approaches described above, rate vs. redshift and rate vs.
age, involve averaging, and hence some loss
of information. In the first approach, one averages over large galaxy
populations, by associating all of the SNe detected at a given
redshift with all of the galaxies of a particular type at that redshift.
\begin{figure}
  \includegraphics[height=.25\textheight]{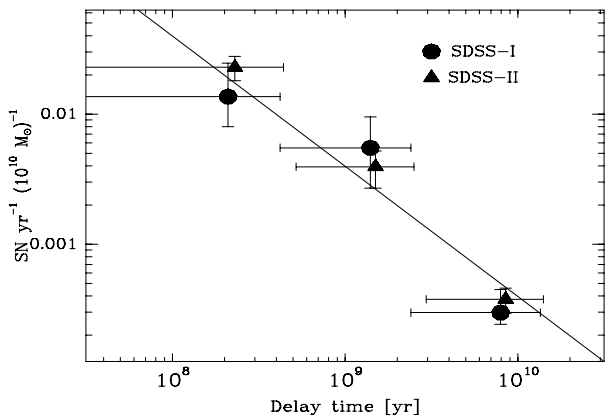}
  \caption{DTDs recovered by comparing SN Ia numbers to indvidual 
galaxy SFHs. Circles are the DTD found by Maoz et al. (2011) 
for
  the Lick Observatory SN Search galaxies and their SNe, that 
have a SDSS-I spectrum with a VESPA (Tojeiro et al. 2009) SFH reconstruction.
 The DTD
  shown uses 49 SNe Ia found among
 1900 galaxies.
Triangles
are the DTD found (Maoz \& Mannucci, in preparation) 
by applying the same inversion algorithm to the sample
of 67,656 galaxies in SDSS-II that have spectra with VESPA SFH reconstructions,
and the 148 SNe Ia that they hosted, similar to the sample analyzed by Brandt et al. (2010). A $t^{-1}$ power law is, again, shown for 
comparison. 
 }
\label{mnm}
\end{figure}
In the second approach, as already noted above, 
a characteristic age for a sample of
galaxies replaces the detailed SFH of the individual galaxies in a SN survey.
Maoz et al. (2011) presented 
a method for recovering the DTD which avoids this averaging.
In the method, the SFH of every individual galaxy, or even every galaxy
subunit, is convolved with a trial universal DTD, and the resulting 
current SN Ia rate is compared to the number of SNe the galaxy
 hosted in the survey
(generally none, sometimes one, rarely more). DTD recovery is treated
as a discretized linear inverse problem, which is solved statistically.
Since the observed numbers of SNe are always very small, Cash (1979)
 statistics  are used.  
Maoz et al. (2011) applied the method to a subsample of the 
LOSS galaxies, and the SNe that they hosted (Leaman et
al. 2011; Li et al. 2011a,b).  From the 15,000 LOSS
survey galaxies, they chose subsamples 
 having spectral-synthesis-based SFH reconstructions  by Tojeiro et
al. (2009), based on spectra from the SDSS. In the recovered DTD
(Figure~\ref{mnm}),  Maoz et al. (2011)
 find a significant detection of both
a prompt SN~Ia component, that explodes within 420~Myr of star formation, 
and a delayed SN~Ia with population that explodes after $>2.4$~Gyr. 

A closely related DTD reconstruction method has been applied by Brandt et
al. (2010) to a different sample, 
the SNe~Ia from the SDSS II survey (Frieman et al. 2008; Sako et
al. 2008), 
conducted by repeatedly imaging 
Stripe 82 of the SDSS. Brandt et al. (2010) also used 
Tojeiro et al. (2009) SFHs, with the same time bins. However, 
rather than directly
fitting the actual number of SNe observed per galaxy, as done by Maoz et al.
(2011), they aimed to 
reproduce the mean spectrum of the SN host galaxies. Like Maoz et
al. (2011), they detected both a prompt and a delayed SN~Ia population.
We have applied the Maoz et al. (2011) algorithm also to an SDSS-II
sample that is similar to the Brandt et al. (2010) sample, but is 
selected somehwat differently and is larger (Maoz \& Mannucci, in
preparation). With this larger sample we detect, at $4\sigma$ significance,
not only a the prompt and delayed components of the DTD, but now
also the intermediate, $0.42<\tau<2.4$~Gyr, component of the SN Ia DTD.
Figure~\ref{mnm} shows together our SDSS-I and SDSS-II 
DTD reconstructions, and their good agreement with a $t^{-1}$ power law.

Brandt et al. (2010) used the ``stretch parameter'', $s$, of the SN
light curves, to divide their SN Ia sample into a ``high-stretch''
subsample and a ``low-stretch'' one, and derived the DTD for
each subsample. They found that luminous, high-stretch,
 SNe Ia tend to have most
of their DTD power at short delays, while low-stretch, underluminous,
SNe Ia have a DTD that peaks in the longest-delay bin. This is the
first derivation of a bivariate DTD, $\Psi(\tau, s)$, albeit with just
three delay-time bins and two stretch bins. (Here, DTD is no longer an
appropriate name, as this is now the bivariate distribution of 
delay times and stretches. A more suitable name, as in other  fields,
would be the bivariate response-, or transfer-, function, e.g. Bentz et
al. 2010). The bivariate SN Ia response function is the thing to aim
for in future surveys, that will have large numbers of well-characterized SNe,
found among samples of galaxies with well-modeled SFHs. The bivariate
response 
contains information
that is additional to the distribution's
 univariate projection, the DTD, as it gives
not only the age of the progenitor systems but also the run of
explosion energies for each progenitor age. 

\subsubsection{SN remnants in nearby galaxies with SFHs based on resolved
  stellar\\ populations}
Another application of the idea to reconstruct the DTD while taking
 into account SFHs, rather than mean ages, was made
by Maoz \& Badenes (2010). They
applied this method to a sample of 77 SN remnants in the Magellanic
Clouds, which were compiled in Badenes, Maoz, \& Draine (2010). 
The Clouds have very detailed SFHs in many small individual spatial
cells, obtained by Zaritsky \& Harris (2004) and Harris 
\& Zaritsky (2009), by fitting model stellar
isochrones to the resolved stellar populations. Thus, one can compare
the SFH of each individual cell to the number of SNe it hosted (or did not)
over the past few kyr, as evidenced by the observed remnants. This
turns the remnants in the Clouds into
an effective SN survey, although several complications need to be
dealt with (see Badenes et al. 2010 and Maoz \& Badenes 2010).  
The SFHs are much more detailed and reliable than
those based on stellar population synthesis of integrated galaxy
spectra.  As there is no way to 
distinguish between core-collapse and Ia SNe in old remnants, a
very early DTD bin, at delays $0<\tau<35$~Myr, is included in 
the reconstruction;
 the signal in that bin is due to the core-collapse SNe. Unfortunately,
since the time-integrated ratio of core-collapse 
SNe to SNe Ia from a stellar population 
is about 5:1 (Maoz et al. 2011), only about a dozen of the Cloud remnants
are from SNe Ia. This small number of remnants, with the attendant large
statistical errors, means that the SN Ia part of the DTD (at $\tau>35$~Myr)
can be binned, at most, 
into two time bins. Nevertheless, Maoz \& Badenes (2010) find 
a significant detection of a prompt (this time $35<\tau<330$~Myr) SN~Ia
component. An upper limit on the DTD level at longer delays is
consistent with the long-delay DTD 
levels measured with other methods. The 
ratio between the rates of prompt and delayed SNe Ia is again
consistent with expectations from $t^{-1}$. This is shown in 
\begin{figure}
  \includegraphics[height=.25\textheight]{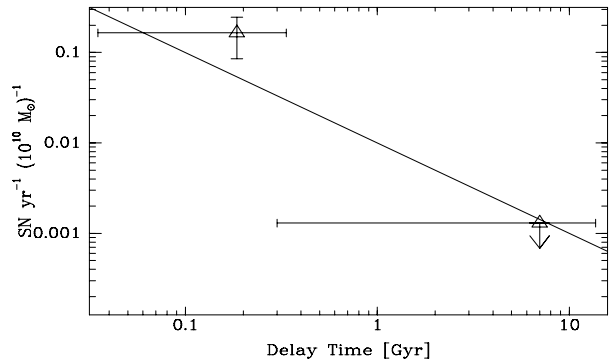}
  \caption{SN Ia DTD from Maoz \& Badenes (2010), based on SN remnants 
in the Magellanic Clouds, compared to SFHs from resolved stellar populations in individual spatial cells, from Zaritsky \& Harris (2004, 2009). A core-collapse
SN bin, at $\tau< 35$~Myr, which is also part of the DTD reconstruction, is not 
shown.  A $t^{-1}$ power law is plotted for 
comparison.
}
\label{lmcfig}
\end{figure}
Fig.~\ref{lmcfig}.
Larger samples can be produced in the future 
via ongoing and proposed deep radio surveys for the
SN remnant
populations in additional nearby galaxies, such as M33 and M31, and
by using 
their spatially differentiated SFHs, 
again based on the resolved stellar populations.

An objection that may arise, when considering this approach, is
that one cannot correctly deduce SN delay times by comparing, on the
one hand, star
formation rates in a small projected piece of a galaxy to, on the
other hand, the SNe that
this region of the galaxy is seen to host, since random velocities cause the SN
progenitor, by the time it explodes, to have drifted far from its
birth location. While this objection is indeed valid if one is comparing 
SN numbers to the mean stellar ages at their locations, it does not apply if, as
here, we are considering detailed SFHs (rather than mean ages), for
full ensembles of galaxy cells and SNe. The reason is that both the
SN progenitors and their entire parent populations undergo the same
spatial diffusion within a galaxy over time. This is explained in more
detail, and with some examples,
in Maoz et al. (2011) and Maoz \& Badenes (2010).

\section{Synthesis}

\subsection{The form of the DTD}
\begin{figure}
  \includegraphics[height=.25\textheight]{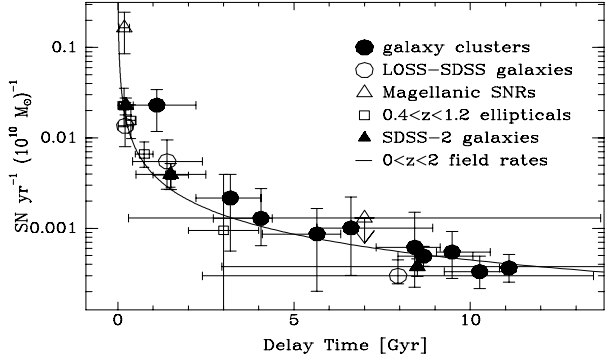}
  \caption{All of the DTDs from previous figures. The solid
curve is a  $t^{-1}$ power law, of the form that gives a good
 fit to the volumetric
SN Ia rates versus redshift (Section~\ref{fieldrates}), and also 
 describes well all of these independent DTD derivations.
}
\label{alldtdlinear}
\end{figure}

\begin{figure}
  \includegraphics[height=.25\textheight]{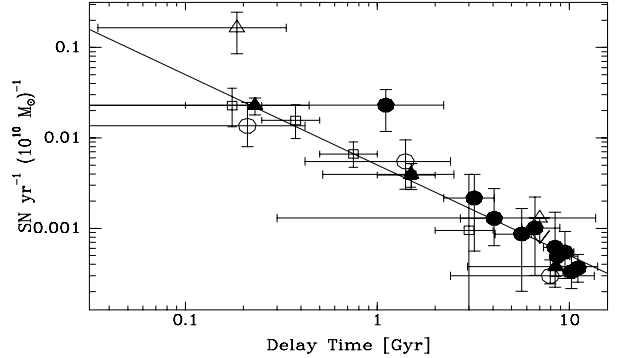}
  \caption{Same as Fig.~\ref{alldtdlinear}, but with a logarithmic
  time axis}.
\label{alldtdlog}
\end{figure}

To synthesize the results reviewed above, 
Figure~\ref{alldtdlinear} shows, on one
plot, the DTD measurements described previously: the DTD based on 
galaxy-cluster SN Ia rates (Maoz et al. 2010); the DTD from the ages
of high-$z$ field ellipticals (Totani et al. 2008); the DTD from the nearby
LOSS galaxies with their SDSS-based SFHs, and the SNe they hosted 
(Maoz et al. 2011); 
the DTD from all SDSS-II galaxies having spectroscopic
SFH reconstructions, and their SNe (Maoz \& Mannucci, in preparation); 
 the DTD
from the Magellanic Cloud SN remnants by Maoz \& Badenes (2010); and 
(solid curve) a $t^{-1}$ 
power-law DTD that provided a good fit the volumetric field SN
rates, when compared to the
cosmic SFH (Graur et al. 2011, see Figure~\ref{fieldrates}). 
Except for this last DTD, all measurements are at the levels that emerge
from the data themselves -- there has been no vertical adjustment
of the points to each other. 
Figure~\ref{alldtdlog} 
shows the same data, but 
on a logarithmic time axis that illustrates more clearly the situation at
short time delays.

The picture emerging from
Figs.~\ref{alldtdlinear}--\ref{alldtdlog}
is remarkable. For one, all of these diverse DTD determinations, based
on different methods, using SNe Ia in different environments and
at different redshifts, agree with each other, both in form and in absolute
level. At delays $t>1$~ Gyr, there seems to be little doubt that the DTD is 
well described by a power law of the form $t^{-s}$, with $s\approx 1$.
At delays $t < 1$~Gyr, the picture is perhaps not as clear cut. Nonetheless, it
{\it is} clear that the DTD does peak in that earliest time bin.
It may continue to rise to short delays with the same slope seen at 
long delays, or it may transit to a shallower or steeper rise, but it certainly
does not fall. The explosion of at least $\sim 1/2$ of SNe Ia 
within 1~Gyr of star formation is, by now, probably an inescapable fact.
The solid curve plotted in Figs.~\ref{alldtdlinear}--\ref{alldtdlog}
is 
\begin{equation}
\Psi(t)=4\times 10^{-13}~{\rm SN~yr}^{-1}M_\odot^{-1}\left(\frac{t}{1~{\rm Gyr}}\right)^{-1}.
\end{equation}
Its integral over time between 40~Myr and 10~Gyr is
$N_{\rm SN}/M_*=2.2\times 10^{-3}~{\rm M_\odot}^{-1}$.


Recalling the generic predictions of theoretical models, described is
Section~\ref{theory}, the observed DTD is strikingly similar to the
simplest expectations from the DD model, namely an approximately
$t^{-1}$ power law extending out to a Hubble time. The SD models, we
recall, though having a rich variety, tend to predict 
no SNe~Ia at delays 
greater than a few Gyr (with the exception of models that succeed in
producing an efficient red-giant donor channel).
At face value, the observed results 
would mean that SD SNe Ia do
not play a role in producing the DTD tail clearly seen at long delays
in the observations. However, the present data cannot exclude also an SD 
contribution at short delays, present in tandem with a DD component that
produces the $\sim t^{-1}$ power law DTD at long delays.

\subsection{The normalization of the DTD}

Apart from the form of the DTD, there is also fairly good agreement 
(though perhaps some tension),
among all
the derivations, on the DTD normalization, or equivalently, its integral
between 40~Myr and
a Hubble time, $N_{\rm SN}/M_*$, i.e., the time-integrated number of 
SNe Ia per stellar mass formed. 
Table~1 summarizes these numbers (as always, with a consistent
assumed IMF -- the Bell et al. 2003 diet-Salpeter IMF that  
simulates a realistic IMF with a low-mass turnover).

Several
 of the normalizations appear nicely consistent
with $N_{\rm SN}/M_*\approx 2$, 
in units of 
$10^{-3}~{\rm M_\odot}^{-1}$. 
One exception to this, on the high side, is
the number based on the iron mass content of  galaxy clusters,
$N_{\rm SN}/M_*> 3.4$. This number, which is based on cluster 
iron abundance, gas fraction, and assumed core-collapse SN iron yield,
sets the lowest-delay bin in the DTD from clusters.
However, 
as seen in Figs.~\ref{alldtdlinear}-\ref{alldtdlog}, 
the other cluster DTD points, which come directly from cluster SN rate 
measurements (rather than from the iron constraint), 
appear to be in good agreement with most of the DTDs from other methods. 
This is seen quantitatively also in Table~1, which gives the best-fit
$N_{\rm SN}/M_*$ normalizations, and $1\sigma$ errors,
 based on the cluster SN rates alone,
{\it assuming} a DTD of the form $t^{-1}$, or   $t^{-0.9}$ (the cluster
rates alone do not constrain well the power-law index).  
This suggests that there may be an error in one or more of the assumptions
of the iron-based point: the iron abundance, or the gas-to-stellar mass ratio 
in clusters may have been systematically
overestimated; or the contribution of core-collapse SNe to cluster iron
enrichment may be underestimated, e.g., if pair-instability SNe are major 
iron suppliers (e.g., Quimby et al. 2011; Kasen et al. 2011). 
It has been pointed out (Bregman et al. 2010)
that the roughly constant iron abundance in galaxy clusters of different 
total masses, despite the large range in their 
gas-to-stellar-mass ratios, 
indicates
a major source of iron that is unrelated to the present-day stellar population
in cluster galaxies. Similar conclusions have been reached from the analysis
of radial abundance gradients in clusters (Million et al. 2011).

Another high $N_{\rm SN}/M_*$ value comes from the SN remnants
in the Magellanic Clouds. Here, it is quite possible that the short-delay bin
is contaminated by core-collapse SNe, and hence is overestimated (see Maoz \& 
Badenes 2010). Furthermore, the overall normalization in this case rests
on the assumption that all stars above $8~M_\odot$ produce core-collapse SNe,
an assumption that may not hold if some fraction of such stars collapse
directly into black holes (e.g. Horiuchi \& Beacom 2011), in which 
case $N_{\rm SN}/M_*$ would be reduced correspondingly.   

Deviating  on the low side, the volumetric field SN Ia rates versus redshift 
suggest $N_{\rm SN}/M_*\approx 1$. This could be an indication that 
most SFH estimates have been overestimated, perhaps due to over-correction
for extinction, by 50\%, or even a factor of 2 (see discussion of this point
in Graur et al. 2011). It is hard to believe that the lower $N_{\rm SN}/M_*$
value implied by the volumetric rates is a real effect due, e.g., to 
environment,
metallicity, or evolution, since some of 
the measurements (e.g. galaxy clusters) that give high  $N_{\rm SN}/M_*$ 
values span redshift ranges that are similar to those of the volumetric 
measurement,
and others (e.g. SDSS-II) were obtained  in 
environments that are similar to those of the volumetric one. 
It thus remains to be seen if the current   
observed range of $N_{\rm SN}/M_*\approx (0.5-3.5)$ will turn out to
indicate a real spread; or to be the result of a universal value that is
somewhere in this range, perhaps 
$N_{\rm SN}/M_*\approx 2$, but that is affected in some cases by 
random and systematic errors. 
\begin{table}[h]
\begin{center}
\caption{DTD normalization results}\label{tableexample}
\begin{tabular}{lcl}
\hline Source & $N_{\rm SN}/M_*$& Ref. \\
              & $[10^{-3}~{\rm M_\odot}^{-1}]$&\\
\hline 
Cluster Fe content& $>3.4$ & a\\
Magellanic SN remnants & $>2.7$& b\\
Cluster rates, $\Psi\propto t^{-1}$&$2.5\pm0.4$&a,c\\
Cluster rates, $\Psi\propto t^{-0.9}$&$2.0\pm0.2$&a,c\\
LOSS SDSS-I galaxies & $2.0\pm 0.6$ & d\\
SDSS-II galaxies & $2.1\pm 0.3$& e\\
Volumetric rates to $z=2$& $1.0\pm 0.5$& f\\
\hline
\end{tabular}
\end{center}
$^a$ Maoz et al. (2010) \\
$^b$ Maoz \& Badenes (2010) \\
$^c$ This work \\
$^d$ Maoz et al. (2011) \\
$^e$ Maoz \& Mannucci, in prep.\\
$^f$ Graur et al. (2011)\\
\end{table}

Compared to these observed DTD normalizations, the theoretical DD 
models do not fare
too well. As already noted by Maoz (2008), Ruiter et
al. (2008), Mennekens et al. (2010), and Maoz et al. (2010), 
binary synthesis DD models underpredict
observed SN rates by factors of at least a few, and likely by
more. One way of alleviating this
inconsistency with the observations would be to include
sub-Chandra mergers in the accounting (see Section~\ref{wdbinaries},
below). Alternatively,
 Thompson (2010) has proposed that
at least some of the SN Ia progenitors may be triple systems,
consisting of a WD-WD inner binary and a tertiary that induces Kozai
(1962) oscillations in the inner binary, driving it to higher
eccentricity and shortening the time until a gravitational-wave-driven
merger
between the two WDs. The possibility of detecting such triple systems
through their gravitational-wave signals is explored by Gould (2011).
Another rate-enhancement scenario is through
 an increase in the number of close
binaries, if most SNe Ia  occur in star clusters. Dynamical 
encounters between binaries and other cluster stars will harden the 
binaries (Shara \& Hurley 2002). The effect has been used to explain 
the enhancement in the number  
of low-mass X-ray binaries observed in globular clusters 
(Sarazin et al. 2003). However, 
Washabaugh \& Bregman (2011) place upper limits on the presence of globular 
clusters at the locations of SNe Ia in elliptical galaxies observed with HST,
thus ruling out globular clusters as a significant global 
SN Ia rate enhancement mechanism.

\section{So, where are those\\ pre-merger WD binaries?}
\label{wdbinaries}
If DD mergers produce SNe Ia, the progenitor systems should be around us.
It is actually easy to estimate quite
accurately what
fraction of local WDs must be SN Ia progenitors, in order to explain 
the SN Ia rate in the context of the DD scenario. From the
LOSS
survey, the SN Ia rate per unit stellar mass 
in large-ish Sbc galaxies is about $1\times 10^{-13}~{\rm
  yr}^{-1}{M_\odot}^{-1}$ (Li et al. 2011b). We live in a typical
region (the disk) of such a galaxy. In the solar neighborhood the
ratio of stellar mass to WD number is $18.5 ~M_\odot/$WD (based on the 
local stellar mass density, $0.085 M_\odot~{\rm pc}^{-3}$, 
McMillan 2011;  and the local WD number density, 
$0.0046~{\rm pc}^{-3}$, Harris et al. 2006). Multiplying the SN Ia rate  
by the stellar-mass to WD ratio, and by a Hubble time, 
 2.5\% of local WDs should be SN Ia progenitors that will merge 
within a Hubble time. (This assumes
a roughly constant star-formation rate in the disk, which would lead
to a constant, steady-state, SN rate.) 
We note that this estimate circumvents the large uncertainties in
the total stellar mass of the Galaxy, its SN rate, and its SFH, 
uncertainties which normally enter such estimates. 
\begin{figure}
  \includegraphics[height=.34\textheight, angle=90]{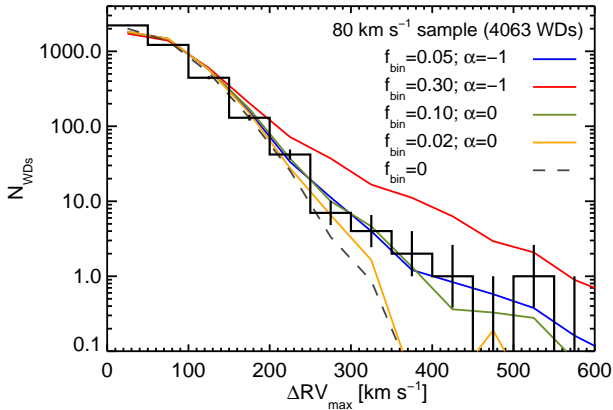}
  \caption{Distribution (histogram) 
of maximum radial velocity differences between 
epochs, among 4063 DA-type WDs from the SWARMS survey, extracted from the
SDSS spectral database. Curves show expectation values from
illustrative models of the WD binary population, in terms of binarity fraction 
and the power-law index of the initial WD separation distribution. Two models
shown fit the data well, two are rejected at high confidence. The dashed
curve is for a model with no binaries, and thus shows the part of the 
distribution that is due solely to velocity errors. 
Acceptable models
turn out to have a local WD merger rate similar to the local SN Ia rate 
per unit stellar mass. 
 }
\label{wdvdist}
\end{figure}

As already noted, SPY
(Napiwotzki et al. 2004; Geier et al. 2007) 
surveyed $\sim 1000$ local WDs and found no binaries that will clearly  
end up as super-Chandra mergers within a Hubble time. According to the 
above estimate, there should have been about 25 such systems,
if the super-Chandra DD scenario is to explain the SN Ia rate.
The efficiency of SPY has not been reported, but it is unlikely to be
so low, and hence this null result argues against this scenario.
On the other hand, one must remember that all observed WD samples are 
flux limited. This might select samples of WDs with binarity fractions
and mass distributions that are distinct from those of the true 
DD progenitor population, which might remain unobserved.

The ongoing SWARMS survey by 
Badenes et al. (2009) is searching for close binaries among
the WDs observed in the SDSS spectral survey. All SDSS spectra were
originally split into  sub-exposures for the purpose of cosmic-ray
rejection. Some subexposures are separated by $\sim 15$~min, sometimes 
by much more, and this permits searching for radial-velocity
variations due to the orbital motions of close DD binaries.
Although the SDSS spectra 
have much lower spectral resolution than SPY (70~km~s$^{-1}$ and 
16~km~s$^{-1}$, respectively), SDSS is a larger sample, and each WD
has, on average, more epochs (increasing the chances to ``catch'' 
a change in radial-velocity).  

Badenes \& Maoz (in preparation) take a statistical approach to
address this question in the SWARMS database. Rather than finding 
all the binaries and characterizing their orbits, they find,
for each  
WD in SWARMS, the 
maximal radial-velocity difference among all its epochs.
They then derive the observed distribution of those maximum velocity 
differences. The distribution probes the parameters 
of the local WD binary population -- binarity fraction, initial separation 
distribution, and mass ratio distribution. To constrain those
parameters, Badenes \& Maoz produce a grid of simulated present-day binary 
populations. These binaries are then ``observed'' with the same
sampling patterns and velocity error distributions as the real data, 
and the simulated maximum-velocity-difference distribution is derived
for each model. The region of binary-parameter space that 
gives velocity difference distributions consistent with the observed
one can thus be found. Furthermore, for every model binary population,
the WD merger rate can be calculated (whether super-Chandra or in general).

Figure~\ref{wdvdist} shows the observed SWARMS maximal radial velocity 
difference distribution, for a sample of 4063 DA-type WDs having 
velocity errors of $<80$~km~s$^{-1}$  per epoch. The smooth curves show
four different WD binary population models, two that reproduce the data
well, and two that are rejected. The observations
thus have the power to discriminate among models. 
 
Intriguingly, the binary population models that do
reproduce the observed velocity-difference distribution have a
super-Chandra WD merger rate that is an order of magnitude too low to 
account for the SN Ia rate, as speculated above based on the SPY null
result. However, the {\it general} WD merger rate (i.e., all merged
masses), is remarkably similar to the SN Ia rate requisite,
$1\times 10^{-13}~{\rm
  yr}^{-1}{M_\odot}^{-1}$.
For some plausible assumptions about the primary and secondary mass
distributions of the WDs, the majority of
those mergers will be similar-mass mergers (e.g., 70\% will have mass
differences less than $0.2 M_\odot$, and 40\% 
 less than $0.1 M_\odot$) and with total masses not too far 
below the Chandrasekhar mass. This raises again the scenario of
sub-Chandra DD mergers as a way of explaining all SNe Ia (e.g. van
Kerkwijk et al. 2010; Guillochon et al. 2010; Ruiters et al. 2011). 
Apart from the the increased numbers,
sub-Chandra merger products have lower central densities. Detonations
at such densities may give SN  ejecta with the correct mix of
iron-peak elements,
intermediate-mass elements, and unburned carbon and oxygen, without 
resorting to the deflagration-delayed-detonation scheme.
 
\section{Additional SN Ia rate\\ phenomenology, explained\\ or not}
\label{ratepuzzles}
In the course of the SN Ia rate studies of the past few years, various
dependences of SN Ia rates on host galaxy properties and environments 
have been seen. We briefly review them here, comment on their current 
observational status, and on whether they are naturally explained in
the context of the recent developments concerning the DTD.

\subsection{Enhanced SN rates\\ in radio galaxies}
Della Valle \& Panagia (2005) and Della Valle et al. (2005), 
analyzing the Cappellaro et al. (1999)
SN sample, found a factor-of-4 enhancement of the SN Ia rate in radio-loud 
early-type galaxies, compared to radio-quiet ones. 
They interpreted this as 
evidence for a population of SNe Ia with a $\sim 100$~Myr delay after star 
formation.
The idea was that an episode of gas accretion or capture of a galaxy
fuels the central black hole, producing the radio luminosity, while
simultaneously triggering star formation. Assuming the lifetime of the
radio phase is   $\sim 100$~Myr, the enhanced SN~Ia rate would be
associated with progenitors of this age in starburst population. 
 An objection to this 
interpretation (Greggio
et al. 2008) was that the same radio galaxies (always early types)
were never seen to host the core-collapse SNe that one would also 
expect from a young starburst. 
Some support for the higher SN Ia rates in radio galaxies,
though not highly significant, has been found in the SNLS sample
by Graham et al. (2010).
The issue should be resolved soon by comparison of the radio-loud 
versus radio-quiet rates 
in the LOSS sample (Li et al. 2011a,b).

\subsection{Enhanced SN rates\\ in galaxy clusters}
There have also been reports of enhanced SN Ia rates in
cluster early-type galaxies, as opposed to field ellipticals.
Mannucci et al. (2008) found such an enhancement in the Cappellaro et
al. (1999) sample, while noting it was only marginally significant.
Recent results by Sand et al. (2011), based on the MENEACS
survey (Sand et al. 2011), give a low-$z$ cluster rate that is
intermediate to the field and cluster elliptical rates of 
Mannucci et al. (2008), but consistent with both to within errors.
Thus, an cluster rate enhancement is not yet rejected, but 
its reality is questionable.

\subsection{The SN rate-size relation}
Most recently, Li et al. (2011b) have discovered a ``rate-size
relation'' in the LOSS data. Among SNe~Ia hosted by specific Hubble
types of galaxies, the rate per unit mass depends on various measures
of host-galaxy ``size'', such as mass or infrared luminosity.
Such an effect is expected in star-forming galaxies, because of the
known anticorrelation between galaxy mass and specific star formation
(e.g., Schiminovich et al. 2007). However,
the effect is seen even in the early-type hosts in LOSS, although its
significance in that case is low.
Following Mannucci et al. (2005), Li et al. (2011b) 
estimated galaxy stellar masses using $B$ and $K$ magnitudes. The 
rate-size relation could be an artifact of the
uncertainties involved in this approach, e.g., due to the effects of the 
mass-age, mass-metallicity relations (Tremonti et al. 2004), or 
the star-formation-rate-mass-metallicity relation 
(Mannucci et al. 2011) relations. 
 Nonetheless, Kistler et al. (2011) 
have shown that the LOSS rate-size relation, even in the early-types,
 can be reproduced at the observed level, based on a $t^{-1}$ DTD,
and the pheonomenon of ``downsizing''. More massive galaxies were,
on average, formed at earlier epochs (e.g., Gallazzi et al. 2005; Pozzetti et al. 2010; Rettura et al. 2011; Kajisawa et al. 2011). 
When observing a massive
early-type galaxy, which is older compared to a less-massive one, we 
are looking further down the tail of the DTD, and therefore measure
a lower SN Ia rate. 

\subsection{The SN-stretch  host-age relation}
As already noted, the relation between host galaxy ``age'' and 
SN Ia luminosity, or stretch, has been known for some time, and
even roughly quantified in the framework of the DTD picture (Brandt et
al. 2010). It is not hard to imagine, at least in principle,
a physical scenario that would lead to such a relation (see,
e.g., Greggio 2010). For example,
in the DD picture, the post-common-envelope WD separation might be a
function of the total WD mass, with more massive binaries having
smaller initial separations than the low-mass pairs. This would
naturally lead to a dependence between delay time and explosion energy. 
It remains to be seen if such a trend is realized in practice, whether 
in models or in observations of WD binary populations.

\section{Sins of SN Rate\\ measurement,\\ and an analogy}
\label{parable}
Measuring and analyzing SN rates is, in principle, straightforward, but
there are pitfalls where some have gone astray. We believe
it may be useful to list the main ones we have encountered in the
literature (without citing the offenders, you know who
you are). Most of these are self-explanatory.

\subsection{The Seven Deadly Sins}
\begin{enumerate}
\item Not using proper detection efficiency simulations.
\item  Not calculating properly the visibility time of your survey.
\item  Using heterogeneous compilations of SNe from surveys with
unknown sensitivities, to estimate rates.
\item  Not stating what are the assumptions about parameters entering your rate
normalization, such as IMF, $H_0$, formed mass vs. existing mass, etc.
\item  Comparing rates without accounting for the effects of different assumed
normalization parameters by different studies.
\item  Ignoring or not presenting the systematic errors in your rates
or analysis, such as uncertainties in extinction, SFH, or host population 
age distribution.
\item  Comparing a posteriori the properties of SN host galaxies to those
of a different "control sample".
\end{enumerate}

The Seventh Sin has been often committed in recent years, but it is
perhaps the one whose
sinfulness is the most subtle to understand. What is wrong, after all,
with this approach? We observe
in detail some specific galaxies that have been seen to host SNe. We
compare some properties of the SN hosts
to those properties as found in some general sample of galaxies that
seems to be matched to the
SN host sample, e.g., in mass, redshift, luminosity, etc. One would
think that any differences we find must then reflect
something about the SN progenitors. For example, if we find an excess
of stars of a given age
in the SN host  galaxies, would not this indicate the age of the
SN progenitor systems when they explode?
Unfortunately, the answer is "maybe", and often "no". The problem is
that, even though we have made an effort
to choose a "good" control sample, it is still a different sample from
the sample that was actually monitored for SNe,
and from which the host galaxies are drawn (and often, we do not even
know what that monitored sample was,
or it might be a heterogeneous compilation of samples with diverse
selection criteria). As such, there is a great
risk that there are some properties that are different in the SN host
sample and in the control sample, but which
have nothing to do with the presence of SNe in the host sample. We
illustrate this with an analogy.

\subsection{The Parable of the Martian\\ Scientists}
Imagine that the Martians have been studying Earth and its inhabitants
for a while. After the less-than-hospitable reception
they received at Roswell in the mid-20th century, they have wisely decided to
stick to remote sensing observations.
They have developed very high resolution techniques, and are
capable of detecting individual humans on Earth. Although they are not yet
aware of the phenomena of growth and aging that characterize
 Earth-bound life, the
Martians have noticed that humans have a range of  properties (e.g., sizes,
speeds) which they quantify with a parameter that we would call ``age''.
(And, in the future, long after the Martian scientists 
{\it will} have understood the
human life cycle, they will continue to refer to elderly people
as ``early-type humans'', and to young people as ``late-type humans'',
just for the sake of confusing their students and themselves!)
  
One research topic of great interest among the Martian Earthonomers
are "baby humans". The Martians have realized that
babies must play some important role in human physiology. The
Martians have developed sensitive techniques to detect
human babies, based on their particular audio-spectral signatures
(which we call cries). Even some amateur Martian Earthonomers
have become quite adept at discovering babies in this way. The value
of baby detections further increased when it was realized
by the Martians that human babies have a narrow distribution of sizes
at birth, with a dispersion of only 14\% (Subramanian et al. 2005).
As such, they serve as excellent "standard rulers" for setting the
scale in the highest-resolution images, and are thus essential for
mapping the Earth. However, the
physical nature of those same babies is not yet understood. Indeed, a fierce
debate continues among the Martians 
on the question of who are the progenitors of the
babies:
Are they other humans, of a certain age? Or are babies perhaps created 
artificially or spontaneously in some process?
Various theoretical progenitor scenarios having been proposed in the
Martian scientific literature. Unfortunately, the birth of babies
invariably occurs
in optically thick structures (which we call hospitals, houses, huts),
making a direct resolution of the question impossible.

The amateur Earthonomers have discovered that they can reap a large harvest of
baby discoveries if they focus their instruments, which can cover
Earth scales of a few hundred km
at a time, on particular regions. For example, an amateur monitoring
the region that we know as Afghanistan might be able to spot a few
thousand new babies in a single
night!  Some professional Martian Earthonomers decide to exploit this
growing database on new babies to address the baby progenitor
question.
Using pointed observations, they measure the human populations of
a sample of individual houses that were reported by the
amateurs to have hosted new babies. Many
 of these houses are in Afghanistan. However, no Martian study exists 
of the human population in Afghanistan as a whole, nor are known the
heterogenous selection criteria and the effective monitoring times
 used by the amateurs. The professionals therefore
compare the properties of the human population in the Afghan baby host houses 
 to a "control sample" for which a 
population study is already available. Among the few available
options, they choose a study that
was done in the region that we call Spain. This seems like a
reasonable choice. Spain and Afghanistan have comparable areas,
geographic latitudes, and population sizes.
The Spanish human population study seems all the more
suited as it gives populations per house, and many of the Afghan
births indeed
take place in houses, as opposed to the larger structures (hospitals)
where baby births often take place in other regions.

 From their comparison of samples, the Martians quickly
discover that the human age distribution in the Afghan baby
host houses is significantly
different from the age distribution in the Spanish control sample.  
First, there is an excess in the
Afghan host houses of humans that are about 20-40 years old, supporting
some previous theoretical speculation that
this is the age of the baby progenitor human population.
However, in the Afghan houses there is an even larger excess of humans
aged 1-15, a population that is quite rare in the Spanish control sample. 
The Martians
promptly conclude that there
is a bimodal baby progenitor distribution: some babies derive from
humans that are 20-40 years old, and some from humans that are 1-15
years old (remember, the Martians are not aware of the process of 
growth and aging, whereby babies become adolescents and then adults). 
The Martians, furthermore,
note another striking difference. The Afghan baby host houses have a
{\it deficit} of humans aged 45-80, compared to the Spanish control
sample. Perhaps, they speculate,
there is a third baby production channel, in which old humans are
transformed into new babies?

The problem, of course, is that despite their best intentions, the
Martians have been comparing a baby host sample and a control sample
that are {\it not} well matched.
Afghanistan has one of the highest rates of births per capita on Earth, 
while Spain has one of the lowest.
Every baby that is born in Afghanistan is likely to have a good number
of siblings living in the same house, and it is they that constitute
the age-1-15 excess population that the Martians are seeing. A
Spanish baby, in contrast, is most likely to have no siblings. And,
sadly, Afghanistan has among
the shortest life expectancies in the world, while Spain has among the
longest, and this is the true reason for the deficit of older people
in the Afghan sample.

Of course, things need not have turned out so incorrectly for the
Martians. They might have chosen for their control sample a region
with an age distribution that is
more similar to that of their host sample. But the only way of being
{\it sure} that their control sample is in fact adequate would have
been to measure the properties of
the {\it same} population that was monitored for babies, and for the whole of
that population (or at least a randomly selected, properly weighted,
subsample of it). Had they done so, 
they would have found that the presence of $\sim$20-40-year-old humans is 
a more-or-less necessary condition for the appearance of a new baby,
be it in a dwelling in Spain or in Afghanistan, while the presence
or absence of the other age groups is not. 
If, furthermore, the Earthonomers
could figure out the exact time periods during which each house was monitored
for babies that could have been detected (the visibility time), 
they would be able to reconstruct
the birth rate as a function of progenitor age -- the baby delay time 
distribution. With carefully measured samples having enough babies
for good statistics, they would be able to gain important insights
about human physiology and society. For example, they would see
that the DTD is shifted to larger delays in Spain compared to
Afghanistan, reflecting the sociological trend in Western societies for
a later child-bearing age.

\section{Conclusions}
In summary, a host of measurements over the past few years have
revealed an increasingly clear picture of the SN Ia DTD. It is well
described by a power law of index $\approx -1$, going out
to a Hubble time. At delays of $< 1$~Gyr, this shape may continue,
or the slope may change somewhat. The time-integrated SN~Ia 
production efficiency is about $2\pm 1$ SN Ia events
 for every $1000 {\rm M_\odot}$
formed in stars, i.e. it is now known to better than a factor of 2. 
(This is often expressed as the fraction of $3-8 M_\odot$ stars that 
eventually explode as SNe Ia; the above range translates to 3-10\%.)
The uncertainties regarding the shape and normalization
of the DTD are dominated 
by the uncertainties in the monitored 
galaxy stellar populations and the SFH.

The observed
DTD form is strikingly similar to the form generically expected, due
to fundamental gravitational wave physics, in the 
DD scenario. The efficiency of SN Ia production by detailed models still
falls short of the observed number, by at least a factor of a few.
The competing SD model makes predictions that differ from the
observations both in DTD form and in the absolute numbers of
SNe. Given the disagreement among the SD calculations themselves, it is
not yet clear if this is a problem of the SD model or of its calculation.
But, keeping all these caveats in mind, the current picture appears
to support the DD model.
In the process, several of the puzzles
that have arisen concerning SN Ia rates, as reviewed above, have been
explained or have disappeared. The local WD population appears to have
an insufficient number of close binaries that will merge within a
Hubble time as super-Chandra objects, according to the classical DD
picture. If however, it is sub-Chandra mergers that produce normal SNe
Ia, there may, in fact, be enough such binaries to reproduce the SN Ia rate.

In terms of the future, several developments are unfolding. Two ongoing
multi-cycle treasury (MCT) programs with HST,
CLASH (Postman et al. 2011) and 
CANDELS (Grogin et al. 2011; Koekemoer et al. 2011) aim to 
measure the SN rate out to $z=2$ and perhaps even somewhat beyond.
This can sharpen our view of the shorter delay times in the DTD, and
test for the influence of other parameters, such as metallicity, on the
rates (e.g., Gallagher et al. 2008; 
Meng \& Yang 2011a; Bravo \& Badenes 2011; Kistler et al. 2011).
Deep radio surveys for SN remnants with the EVLA will produce large
samples of remnants in additional Local Group galaxies, beyond the
Magellanic Clouds. Ongoing high-resolution imaging of
those same galaxies with HST will produce the data for region-by-region
SFHs based on resolved stellar populations. With these two datsets 
combined, it will be possible to reconstruct the DTD at short delays,
where the greatest uncertainty remains observationally, and where an
SD contribution may still play a role. Finally, larger samples of SNe and host
galaxies, e.g. from the upcoming Dark Energy 
Survey\footnote{\tt http://www.darkenergysurvey.org}
 and the 
HyperSuprimeCam Survey with 
Subaru \footnote{\tt http://www.astro.princeton.edu/$\sim$rhl/HSC}
will permit addressing the
bivariate SN Ia response function -- the next dimension beyond the
DTD, which will start to connect between the progenitors and the observed
features of the explosions themselves.

\section*{Acknowledgments} 
We thank Gijs Nelemans and the referee, Laura Greggio, for valuable comments.
DM acknowledges support by the Israel Science Foundation.
FM acknowledges partial financial support
of the Italian Space Agency, through contracts ASI-INAF
I/016/07/0 and I/009/10/0, and PRIN-INAF 2008.
FM also acknowledges support by NASA through a grant from the Space 
Telescope Science Institute, which is operated by the Association of 
Universities for Research in Astronomy, Incorporated, under NASA 
contract NAS5-26555.

\newcommand\aj{{AJ}}%
\newcommand\araa{{ARA\&A}}%
\newcommand\apj{{ApJ}}%
\newcommand\apjl{{ApJ}}%
\newcommand\apjs{{ApJS}}%
\newcommand\ao{{Appl.~Opt.}}%
\newcommand\apss{{Ap\&SS}}%
\newcommand\aap{{A\&A}}%
\newcommand\aapr{{A\&A~Rev.}}%
\newcommand\aaps{{A\&AS}}%
\newcommand\azh{{AZh}}%
\newcommand\baas{{BAAS}}%
\newcommand\jrasc{{JRASC}}%
\newcommand\memras{{MmRAS}}%
\newcommand\mnras{{MNRAS}}%
\newcommand\na{{New Astronomy}}%
\newcommand\pra{{Phys.~Rev.~A}}%
\newcommand\prb{{Phys.~Rev.~B}}%
\newcommand\prc{{Phys.~Rev.~C}}%
\newcommand\prd{{Phys.~Rev.~D}}%
\newcommand\pre{{Phys.~Rev.~E}}%
\newcommand\prl{{Phys.~Rev.~Lett.}}%
\newcommand\pasp{{PASP}}%
\newcommand\pasj{{PASJ}}%
\newcommand\qjras{{QJRAS}}%
\newcommand\skytel{{S\&T}}%
\newcommand\solphys{{Sol.~Phys.}}%
\newcommand\sovast{{Soviet~Ast.}}%
\newcommand\ssr{{Space~Sci.~Rev.}}%
\newcommand\zap{{ZAp}}%
\newcommand\nat{{Nature}}%
\newcommand\iaucirc{{IAU~Circ.}}%
\newcommand\aplett{{Astrophys.~Lett.}}%
\newcommand\apspr{{Astrophys.~Space~Phys.~Res.}}%
\newcommand\bain{{Bull.~Astron.~Inst.~Netherlands}}%
\newcommand\fcp{{Fund.~Cosmic~Phys.}}%
\newcommand\gca{{Geochim.~Cosmochim.~Acta}}%
\newcommand\grl{{Geophys.~Res.~Lett.}}%
\newcommand\jcp{{J.~Chem.~Phys.}}%
\newcommand\jgr{{J.~Geophys.~Res.}}%
\newcommand\jqsrt{{J.~Quant.~Spec.~Radiat.~Transf.}}%
\newcommand\memsai{{Mem.~Soc.~Astron.~Italiana}}%
\newcommand\nphysa{{Nucl.~Phys.~A}}%
\newcommand\physrep{{Phys.~Rep.}}%
\newcommand\physscr{{Phys.~Scr}}%
\newcommand\planss{{Planet.~Space~Sci.}}%
\newcommand\procspie{{Proc.~SPIE}}%


\end{document}